# HOMs Simulation and Measurement Results of IHEP02 Cavity*

ZHENG Hong-Juan (郑洪娟)[1] ZHAI Ji-Yuan(翟纪元)[1] ZHAO Tong-Xian(赵同宪)[1] GAO Jie (高杰)[1]

1 (Key Laboratory of Particle Acceleration Physics and Technology, Institute of High Energy Physics, CAS, Beijing 100049, China)

**Abstract:** In cavities, there exists not only the fundamental mode which is used to accelerate the beam but also higher order modes (HOMs). The higher order modes excited by beam can seriously affect beam quality, especially for the higher $R/Q$ modes. This paper reports on measured results of higher order modes in the IHEP02 1.3GHz low-loss 9-cell superconducting cavity. Using different methods, $Q_e$ of the dangerous modes passbands have been got. The results are compared with TESLA cavity results. $R/Q$ of the first three passbands have also been got by simulation and compared with the results of TESLA cavity.

## 1 Introduction

The layout of IHEP02 1.3GHz low-loss 9-cell superconducting cavity [1] is shown in Fig.1. In order to damp HOMs, two HOM couplers were mounted respectively at the upstream and downstream beam tube. Distances from the HOM couplers to end cells were 65mm and 50mm, which were different from TESLA cavity [2]. Length of upstream beam tube was also different from downstream.

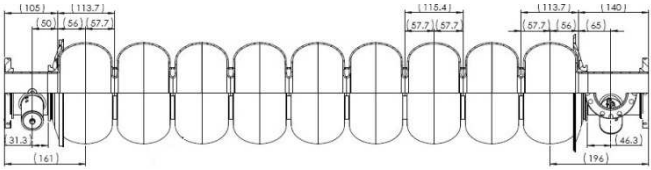

Figure 1: Layout of IHEP02 9-cell cavity

Damping of HOMs in ILC linear collider is necessary. Parasitic modes excited by the accelerated beam may lead to loss of beam quality and additional power dissipation. For HOMs with frequency lower than cut off frequency of the beam tube, their power loss must be extracted by external load. When the moving relativistic particles encounter geometric variations along the structure, such as RF accelerating cavities, vacuum bellows, and beam diagnostic chambers, there are wakefields after the particles go through [3]. Long range wakefields are important for the beam dynamics of a long train of bunches in linear accelerator since this wakefields can cause multibunch instabilities. The long range wakepotential can be represented as a sum over contributions from HOMs. Transverse long range wakefield mainly consider the influence of dipole modes, which requires the $Q_e$ of the HOMs less than $10^5$ [4]. In this paper, we present the simulation results of $R/Q$ and measured $Q_e$ of HOMs.

## 2 Simulation results of $R/Q$

The ratio $R/Q$ is a very important quantity related to the interaction of the beam and cavity. The higher value of $R/Q$ stands for larger energy change between beam and cavity. For monopole HOMs, beam energy loss will be more. For dipole modes, the transverse displacement of the beam will be larger, which will causes emittance larger and even worse it will leads beam loss. It becomes very important to measure modes with large $R/Q$.

An important parameter describing the transverse beam-cavity interaction is transverse shunt impedance. From the Panofsky-Wenzel theorem [3], the transverse momentum change of the particle passing through a cavity excited in a single mode is proportional to a parameter $(R/Q)_\perp$. In this paper, for dipole modes, $(R/Q)_\perp$ is defined as:

$$(\frac{R}{Q})_\perp = \frac{\left|\int E_z e^{j\omega z/c} dz\right|^2}{k^2 r^2 \omega U} = \frac{V_{//}^2}{k^2 r^2 \omega U} \qquad (1)$$

The unit of $(R/Q)_\perp$ is ohms. $k$ is the wave number. $V_{//}$ is the accelerating voltage of a cavity which is defined as:

$$V_{//} = \int_0^d E_z e^{j\omega z/c} dz \qquad (2)$$

$d$ is the length of cavity. In this paper, we define a new $(R/Q)_\perp$ called $(R/Q)'_\perp$ as:

$$(\frac{R}{Q})'_\perp = \frac{V_{//}^2}{r^2 \omega U} \qquad (3)$$

The unit of $(R/Q)'_\perp$ is $\Omega/cm^2$.

Energy generated by HOMs can be coupled by HOM couplers and absorbed by external load. HOMs energy loss can be divided into two parts, power dissipated in the cavity walls $P_0$ and power coupled by the external circuit $P_e$. $U$ is the energy stored in the cavity. It is better to get small $Q_e$ in order to damp the HOMs enough.

*Work supported by the Innovation Program of IHEP, CAS
1) E-mail: zhenghj@ihep.ac.cn

$$Q_e = \frac{\omega U}{P_e} \quad (4)$$

Dangerous modes' $Q_e$ can be got by measuring the microwave parameters of 1.3GHz 9-cell cavity. The measured results affect how to optimize HOM coupler structure in the future.

*R/Q* of the first three HOMs passbands have been simulated. The results of *R/Q* compared with TESLA cavity's results are shown in Fig. 2. In Fig. 2, modes in the box represent the most dangerous three modes which have largest *R/Q* in that passband. The most dangerous modes need to be measured carefully. From the result, we can see that the dangerous modes in IHEP02 1.3GHz low-loss cavity are almost the same with TESLA cavity, except for TM011 mode. This difference is mainly because the different design of these two kind cavities.

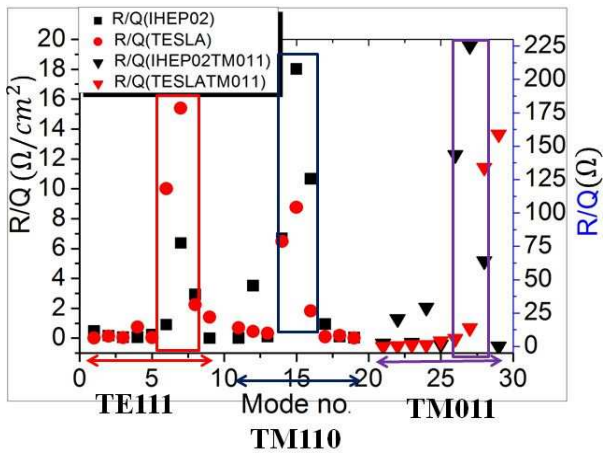

Figure 2: HOMs' *R/Q*.

## 3 Structure of HOM couplers

There are three major varieties of HOM couplers [3]: waveguide, coaxial, and beam tube. Consider the purpose of 1.3GHz superconducting cavity, it requires the connection between cavities short. The coaxial type was chosen, because of their more compact size [4]. The coupler used for IHEP02 1.3GHz low-loss 9-cell superconducting cavity is shown in Fig. 3. According to the rules of excitation, frequencies of modes TE111, TM110 and TM011 are all under the cut-off frequency of beam tube whose radius is 40mm.

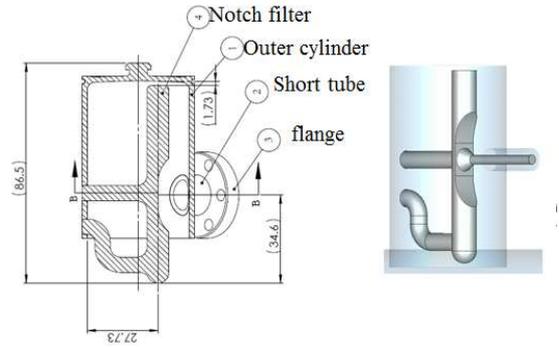

Figure 3: Profile and model of HOM couplers.

In order to sufficiently attenuate the HOMs, there were two HOM couplers mounted respectively on the upstream and downstream [2]. The position of the HOM coupler was 65mm from the edge of the end cell on the upstream and 50mm from the edge of the end cell on the downstream. The design value of the insertion length of the coupling loop tip was 30.25mm from the beam axis. The mounted angle of the upstream coupler and downstream coupler are shown in Fig. 4.

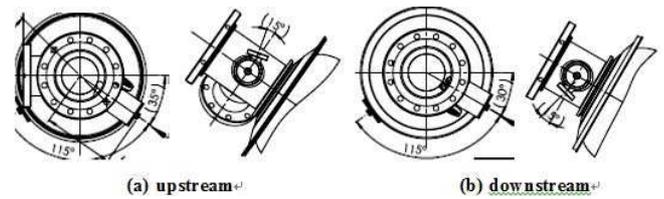

Figure 4: (a) shows the upstream coupler's position and (b) shows the downstream coupler's position

## 4 HOMs measurements

The characteristic of HOMs were measured by using a network analyzer. The ports of the two HOM couplers were used as the excited input and output ports. Measured results of HOMs peak frequencies and the *Q* were performed at room temperature. Measuring device schematic diagram is shown in Fig. 5.

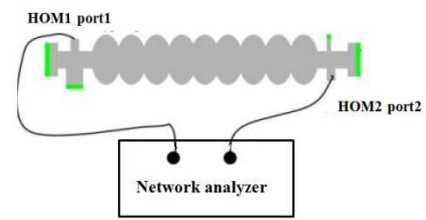

Figure 5: Schematic diagram of the HOM's measurement

### 4.1 Measurements of IHEP02 cavity's HOMs frequencies and passbands

The measured results of frequencies and passbands are shown in Fig. 6. The most dangerous modes were recognized.

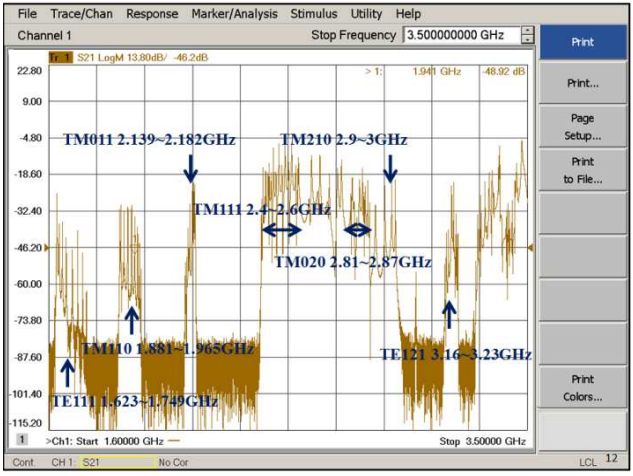

Figure 6. IHEP02 cavity's HOMs frequencies and passbands

### 4.2 Measurements of IHEP02 cavity's HOMs $Q_e$

The excitation signal was imported to the cavity by upstream HOM coupler and the output signal was extracted by downstream HOM coupler. The measured microwave parameters were received from the network analyzer. Bead pulling method [3] was used to measure the field profile.

$Q_e$ of the first three modes with frequencies under the beam tube cut-off frequency were measured. $Q_e$ of the mode TM111 was also measured even if its frequency was above the beam tube cut-off frequency. There was a mode during TM111 passband with the largest $R/Q$ not only of the modes of this passband but also of all calculated modes [5] [6].

Three methods were used to measure $Q_e$, impedance method [7] [8], reflection method [9] and transmission method [10]. Reflection method has a limitation that the coupling coefficient was not too small. Transmission method was used for one port weak coupling. The calculations for these methods are as follows.

a) Impedance method.

The normalized impedance for an equivalent circuit is

$$\overline{Z} = \frac{1}{\frac{1}{\overline{r}} + j(\omega\overline{C} - \frac{1}{\omega\overline{L}})} = \frac{1}{\overline{g} + j\overline{b}} \quad (5)$$

where $\overline{r}$ is equivalent resistance, $\overline{C}$ is equivalent capacitance, $\overline{L}$ is equivalent inductance.

The unloaded quality factor is

$$Q_0 = \frac{\omega_0 W_0}{P_c} \quad (6)$$

where $\omega_0$ is resonance frequency, $W_0$ is energy stored in cavity, $P_c$ is power dissipated in cavity at resonance.

The external quality factor is

$$Q_e = \frac{\omega_0 W_0}{P_e} \quad (7)$$

where $\omega_0$ is resonance frequency, $W_0$ is energy stored in cavity, $P_e$ is power dissipated in external circuit.

Define half power points:

$$P_{c1,2} = 1/2 P_c \quad (8)$$

$$P_{e1,2} = 1/2 P_e \quad (9)$$

from (5)~(9), we can get

$$\overline{b}_{1,20} = \pm\overline{g} \quad (10)$$

$$\overline{b}_{1,2e} = \pm 1 \quad (11)$$

Reflection coefficient,

$$\Gamma = \frac{\overline{z} - 1}{\overline{z} + 1} \quad (12)$$

Voltage standing wave ratio (VSWR),

$$S = \frac{1 + |\Gamma|}{1 - |\Gamma|} \quad (13)$$

from (5)(12)(13), the resonant VSWR is

$$S_0 = \overline{r}(r > 1) \quad (14)$$

$$S_0 = \frac{1}{\overline{r}}(r < 1) \quad (15)$$

from (5)~(15), VSWR at half power points of $Q_0$ and $Q_e$ are

$$(S_{1/2})_0 = \frac{2 + \beta^2 + \sqrt{4 + \beta^4}}{2\beta} ((S_{1/2})_0 < S_0) \quad (16)$$

$$(S_{1/2})_e = \frac{1 + 2\beta^2 + \sqrt{1 + 4\beta^4}}{2\beta} \quad (17)$$

In some cases, the resonant VSWR is very large that it cannot easy to measure. In other cases, the VSWR curve is asymmetric. We can solve these problems by measure arbitrary VSWR value. First, we get the resonant VSWR and frequency by VSWR curve. Then, find arbitrary VSWR value near resonant VSWR. The calculations are as follows.

$$Q_0 = \frac{\omega_0}{\Delta\omega}\sqrt{\frac{(S_x - S_0)(S_x S_0 - 1)}{S_x}}(\beta > 1) \quad (18)$$

$$Q_0 = \frac{\omega_0}{\Delta\omega}\sqrt{\frac{(\frac{S_x}{S_0} - 1)(S_x - \frac{1}{S_0})}{S_x}}(\beta < 1) \quad (19)$$

$$\beta = S_0 (\beta > 1) \quad (20)$$

$$\beta = \frac{1}{S_0}(\beta < 1) \quad (21)$$

$$\frac{Q_0}{\beta} = Q_e \quad (22)$$

$$\frac{Q_0}{1+\beta} = Q_L \quad (23)$$

where $S_x$ is arbitrary VSWR, $\beta$ is coupling coefficient.

b) Reflection method.

Measure the resonant frequency $\omega_0$, loaded quality factor $Q_L$, reflection parameters $S_{11}$(dB) and $S_{22}$(dB), then

$$Q_0 = (1+\beta_1+\beta_2)Q_L \quad (24)$$

There are three different cases for different coupling coefficient.

- Port 1 and port2 are both under coupling:

$$\beta_1 = \frac{1-10^{\frac{S_{11}}{20}}}{10^{\frac{S_{11}}{20}}+10^{\frac{S_{22}}{20}}} \quad (25)$$

$$\beta_2 = \frac{1-10^{\frac{S_{22}}{20}}}{10^{\frac{S_{11}}{20}}+10^{\frac{S_{22}}{20}}} \quad (26)$$

- Port 1 is under coupling and port2 is over coupling. Port 1 and port2 are both over coupling, however $\beta_2 > \beta_1 + 1$

$$\beta_1 = \frac{1-10^{\frac{S_{11}}{20}}}{10^{\frac{S_{11}}{20}}-10^{\frac{S_{22}}{20}}} \quad (27)$$

$$\beta_2 = \frac{1+10^{\frac{S_{22}}{20}}}{10^{\frac{S_{11}}{20}}-10^{\frac{S_{22}}{20}}} \quad (28)$$

- Port 1 is over coupling and port2 is under coupling. Port 1 and port2 are both over coupling, however $\beta_1 > \beta_2 + 1$

$$\beta_1 = \frac{1+10^{\frac{S_{11}}{20}}}{10^{\frac{S_{11}}{20}}-10^{\frac{S_{22}}{20}}} \quad (29)$$

$$\beta_2 = \frac{1-10^{\frac{S_{22}}{20}}}{10^{\frac{S_{11}}{20}}-10^{\frac{S_{22}}{20}}} \quad (30)$$

c) Transmission method.

Measure the resonant frequency $\omega_0$, loaded quality factor $Q_L$, reflection and transmission parameters $S_{11}$(dB) and $S_{21}$(dB), then

$$Q_0 = (1+\beta_1+\beta_2)Q_L \quad (31)$$

$$P_r = P_{in}\Gamma^2 = P_{in}10^{\frac{S_{11}}{10}} \quad (32)$$

$$P_t = P_{in}T^2 = P_{in}10^{\frac{S_{21}}{10}} \quad (33)$$

$$P_{in} = P_r + P_t + P_0 \quad (34)$$

There are two different cases for different coupling coefficient.

- Port1 is under coupling.

$$\beta_1 = \frac{(1-10^{\frac{S_{11}}{20}})^2}{1-10^{\frac{S_{11}}{10}}-10^{\frac{S_{21}}{10}}} \quad (35)$$

$$\beta_2 = \frac{10^{\frac{S_{21}}{10}}}{1-10^{\frac{S_{11}}{10}}-10^{\frac{S_{21}}{10}}} \quad (36)$$

- Port1 is over coupling.

$$\beta_1 = \frac{(1+10^{\frac{S_{11}}{20}})^2}{1-10^{\frac{S_{11}}{10}}-10^{\frac{S_{21}}{10}}} \quad (37)$$

$$\beta_2 = \frac{10^{\frac{S_{21}}{10}}}{1-10^{\frac{S_{11}}{10}}-10^{\frac{S_{21}}{10}}} \quad (38)$$

The measured results of $Q_e$ using these three methods compared with TESLA design goal are shown in Fig. 7. Modes in the box are the most dangerous modes in that passband. Although frequency of TM111 mode is above the cut off frequency of the beam tube, it has two dangerous modes [5] [6]. TM111 mode also need measured. The measurement results show a good damping with all the modes under $10^5$. The lengths from HOM coupler to end cells of IHEP02 cavity were 65mm and 55mm while 45mm for TESLA cavity. From equation $Q_e = \frac{\omega U}{P_e} \propto \frac{1}{E^2} \propto e^{2\alpha z}$ [11], one can got a conclusion that the longer the length, the smaller value of $Q_e$. The damping of TM110 mode in IHEP02 cavity is better than TESLA cavity. $Q_e$ are lower than TESLA BBU limit, this mainly because the different cavity designs. These measured results give a good instruction for the next HOM coupler optimal design.

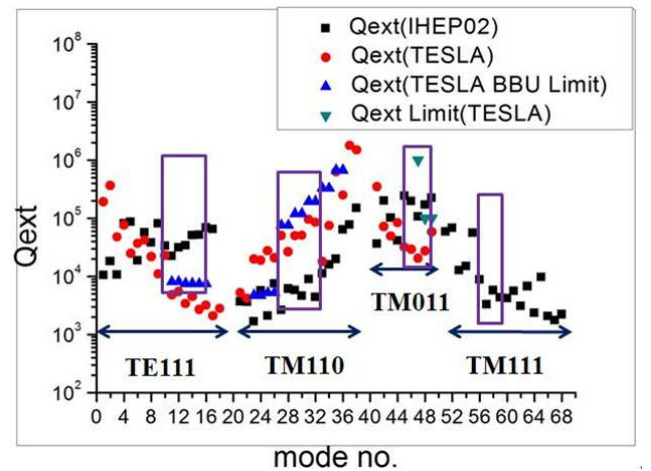

Figure 7: $Q_e$ of IHEP02 cavity compared with TESLA results

In order to get the relationship between $Q_e$ and feedthrough gap, $Q_e$ values with different gaps were also measured. The fifth mode of TM110 (5Pi/9) was choose to measure. Design value of feedthrough gap was 0.5mm. Definition of feedthrough gap is shown in Fig. 8. The measured values are shown in Fig. 9. These measured results give a good instruction for the HOM coupler's feedthrough gap design.

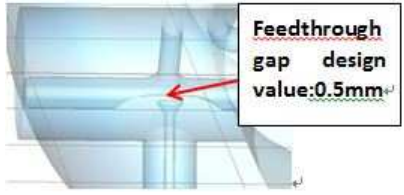

Figure 8: Definition of feedthrough gap

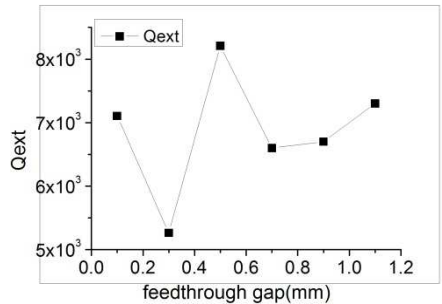

Figure 9: Measurement results of $Q_e$ ~ feedthrough gap

## 5 Conclusion

In this paper, different methods of measurement of $Q_e$ and the measurement results are presented. $(R/Q)_\perp$ simulated results are also got for the first three HOMs passbands. The measurement results show that:

- Very good damping for the dipole passbands TM110 and TM111, $Q_e$ of these modes are all under TESLA BBU limit.
- $Q_e$ values of the monopole passband TM011 are near the $Q_e$ limit of TESLA.
- Damping of first dipole passband TE111 should be improved to keep $Q_e$ below the TESLA BBU limit.

The measurement results provide a good guidance for the optimization of HOM couplers design in the future.

### Acknowledgments

The authors would like to thank the support of RF group of IHEP. We also thank L. S. Huang and S. K. Tian for their advice of learning software.